\documentclass[]{emulateapj}
\usepackage{amsmath}
\usepackage{natbib}
\usepackage{graphicx}
\usepackage{color}
\usepackage{nicefrac}
\usepackage{txfonts}
\usepackage{listings}
\usepackage{color}
\usepackage{cleveref}[2012/02/15]
\usepackage{subfigure}
\usepackage{xspace}
\usepackage{hyperref}
\usepackage{breakurl}
\usepackage{amssymb}
\usepackage{array}
\newcolumntype{L}[1]{>{\raggedright{}}p{#1}}

\usepackage[percent]{overpic}

\newcommand{\git}{\texttt{git}}

\usepackage{color}
\definecolor{gray}{rgb}{0.4,0.4,0.4}
\definecolor{cyan}{rgb}{0.3,0.3,0.3}

\lstdefinelanguage{XML}
{
  morestring=[b]",
  morestring=[s]{>}{<},
  morecomment=[s]{<?}{?>},
  stringstyle=\color{gray}\ttfamily,
  identifierstyle=\ttfamily\bfseries,
  keywordstyle=\color{cyan}\ttfamily\bfseries,
  numberstyle=\color{gray}\ttfamily,  
  numbersep=5pt,             
  morekeywords={attributename,units}
}

\def \aap{A\&A}

\def \apjl{ApJ}
\def \apjs{ApJS}
\def \apj{ApJ}

\def \mnras{MNRAS}

\def \pasp{PASP}

\newcommand{\xml}[1]{{\ttfamily<#1>}}

\slugcomment{\today{}}

\begin{document}

\title{ A proposal for community driven and decentralized astronomical databases \\
    	and the Open Exoplanet Catalogue}

\author{Hanno Rein}
\affil{Institute for Advanced Study, Princeton, NJ 08540}

\begin{abstract}
I present a new kind of astronomical database based on small text files and a distributed version control system.
This encourages the community to work collaboratively.
It creates a decentralized, completely open and democratic way of managing small to medium sized heterogeneous astronomical databases and catalogues.
The use of the XML file format allows an easy to parse and read, yet dynamic and extendable database structure.\vspace{5pt}

The Open Exoplanet Catalogue is based on these principles and presented as an example.
It is a catalogue of all discovered extra-solar planets.
It is the only catalogue that can correctly represent the orbital structure of planets in arbitrary binary, triple and quadruple star systems, as well as orphan planets.
\end{abstract}

\section{Introduction}
Astronomical catalogues and databases are almost as old as astronomy itself. 
Lists of star names and constellations were compiled by many ancient civilizations including the Chinese, Persians, Babylonians, Arabs and Greeks.
Our technology has changed dramatically since then.
However, the basic principles of how these catalogues and databases are compiled and maintained has not changed much at all. 
This proposal presents a new way to manage astronomical databases using modern technology.

In recent years, the internet allowed the open source community to make extensive use of distributed version control systems.
Almost all of the software that we use every day was developed using these systems.
This includes astrophysical projects like the \texttt{pencil} code\footnote{\url{http://www.nordita.org/pencil-code/}}.
But perhaps the most prominent example is the Linux kernel. 
The Linux kernel has over 15 million lines of code and an average of 3500 lines are added each day\footnote{Statistic adopted from \url{http://royal.pingdom.com/2012/04/16/linux-kernel-development-numbers/}.}.
The number of active contributors varies a lot from project to project.
Most projects involve only a handful of developers.
However, over 1000 developers contribute to a new release of the Linux kernel\footnote{Number taken from \url{http://lwn.net/Articles/395961/}.}.
Linus Torvalds is the official maintainer of the project and releases each final version. 
However, his contributions to the kernel itself do not even give him a place in the top 30.

At first sight, it may seem surprising that such a large group of individuals can work together to achieve a common goal.
It might be even more surprising to claim that a similar approach could benefit astronomy.
But there are many similarities between the astronomy and open source community.
We are both dedicated to keeping our results as open and easily accessible as possible.
We both care about security, reliability and credibility.
We are both large communities with many people having diverse backgrounds and needs.
I therefore propose to adopt parts of this workflow to astronomical databases and catalogues.

In the following I will list the main advantages of this approach. 
The two main ideas are using human-readable XML files to store the data and using \texttt{git} as a version control system to manage the database.
I discuss technical details regarding those two ideas in Sections \ref{sec:xml} and \ref{sec:git}.
Although I mention specific technologies, the ideas are very general and do not rely on either XML or \git{}.
Finally, I will present one example of such a database that I have already been implementing, the Open Exoplanet Catalogue.

\section{Main advantages} \label{sec:adv}
There are many advantages to using an open, decentralized approach when managing astronomical databases and catalogues.
In my opinion, the main advantages are the following:
\begin{description}

\item[Community driven]
Anyone can contribute data to the catalogue. 
This includes the original observer who collected the data, other astronomers who spot a typo or even interested members of the general public who might want to help\footnote{One cannot underestimate the enthusiasm of people to do something like this. Just look at wikipedia and the citizen science project zooniverse.}.

\item[Democratic principles]
The maintainer of the database is responsible for including updates into the main database.
This person checks the validity of the submissions and makes sure only those that he thinks are credible are added.

However, sometimes a person might disagree with the maintainer's decisions.
That person can then simply fork\footnote{A fork is a copy of the database. It also copies the entire history of the database.} the project. 
This allows him to stay up-to-date with the latest updates to the main database.
But at the same time he can add his own data which he thinks is credible but the original maintainer disagrees. 
If the maintainer later decides that it was indeed a credible addition, he can just merge the new fork back into the main database.

This is an important point. Let me explain this with a analogy to the Linux kernel. 
In that case most people trust Linus Torvalds and will use his version (the \textit{official} version).
They do not know if they can trust the other maintainers of the more than 1000 forks of the Linux kernel that are available on github\footnote{Github is a free hosting platform for projects based on the \texttt{git} version control system. It is used by many open source projects including the Linux kernel.}. 
However, in certain cases you might still want to have a look at one of these forks.
For example when you want to run Linux on a new Laptop that needs a new driver which has not been included in Torvalds' version yet.
A very similar workflow might occur in astronomy, for example when a new extra-solar planet has been announced but has not been confirmed yet. 

\item[Decentralized structure]
The concept of forks and the resulting decentralized structure have many more advantages.
Suppose the original maintainer stops maintaining the database because he has left the field, is on holiday or simply doesn't have time anymore.
Other members of the community that rely on this database can easily continue updating it without anyone having to formally give up control over the original version. 

Everyone can copy the full database including the entire history of changes to a local machine. 
One can analyze, edit, delete and add data without an internet connection.
There is no need to host and maintain a server (of course this is still possible).

The decentralized structure is furthermore a natural way to back up data.
Let us once again take the Linux kernel as an example. 
The project has probably been forked by millions of people.
No matter how many servers fail or get hacked, there will always be one copy that is still intact. 
Furthermore, it is even possible to continue working together on the project while a major server is not accessible.

\item[Openness]
Anyone can see the entire database structure and all available data. 
There is no web interface required that would hide important parts of the data.
This also speeds up the development of the database because it does not involve programming a user interface.
Of course a web interface can always be implemented at a later time to allow simplified access to the data.

\item[Correctness and privacy]
Many catalogues contain data from several different groups.
Using a version control system to manage the database allows anyone to contribute their own data and make corrections to existing data.
This reduces the number of errors compared to catalogues which are compiled by a single third party.

Observers can enter and verify new and confidential data on a local copy of the database.
The update can be pushed to the main repository at the same time an announcement is made or an embargo lifted. 

\item[Easily extendable]
Separating a database into small text files makes reading and maintaining it very easy. 
If the database in written in the XML file format, adding a new field to the database is trivial.
One only needs to change those files for which new data exists. 
All the other files remain unchanged.
Any existing script that parses the database will be fully backwards compatible without any additional work.

\item[Keeping track of history]
The entire history of each change to the database is recorded in the version control system.
This allows one to go back and redo an analysis with only that part of the data that was available at an earlier time.
One can also search for trends in the data.
For example, one can easily verify that the reported eccentricity for a given planetary systems is almost certainly decreasing as a function of time because new data becomes available and partly counteracts a strong bias in fitting Keplerian orbits \citep{ShenTurner2008}.
Studies like this are not possible if only the current version of a catalogue are available.

\item[Blame and Credit]
Each line in each file can be traced back to the person who made changes to it and to the peer-reviewd publication where the data is taken from. 
Therefore the entry database is verifiable. 
If found that a person made dubious entries across many different files, this can be easily corrected by reverting all those commits to the version control system.
\end{description}

\section{Human-readable XML files} \label{sec:xml}
It is important to stress that the discussion in this paper is very general and does not depend on the specific file format.
There are many options and the choice is often just a matter of personal preference.
There are only two requirements for a file format to work nicely with a version control system: the files have to be text files and each value should be placed in a separate line.
I chose XML for the Open Exoplanet Catalogue because of the flexible tree structure that allows me to construct arbitrary star systems and because of the vast number of parsing libraries that are readily available for many different languages. 

\subsection{Basic features of the XML file format}
\begin{figure}
\caption{Example of a database entry in the Open Exoplanet Catalogue: CoRoT-5.}
\vspace{5pt}
\begin{lstlisting}[language=XML, numbers=left,frame=tb]
<name>CoRoT-5</name>
<rightascension>06 45 07</rightascension>
<declination>+00 48 55</declination>
<distance>400</distance>
<star>
	<mass>1.03</mass>
	<radius>1.186</radius>
	<temperature />
	<planet>
		<name>CoRoT-5 b</name>
		<list>Confirmed planets</list>
		<mass>0.467</mass>
		<radius>1.388</radius>
		<period>4.0378962</period>
		<semimajoraxis>0.0494&\newline&</semimajoraxis>
		<eccentricity>0.09</eccentricity>
		<inclination>85.83</inclination>
		<description>CoRoT-5 b orbits an F-type star in the constellation Monoceros.&\newline&</description>
		<discoverymethod>transit&\newline&</discoverymethod>
		<lastupdate>09/07/18</lastupdate>
		<discoveryyear>2008&\newline&</discoveryyear>
	</planet>
</star>
\end{lstlisting}
\vspace{5pt}
\label{lst:corot}
\end{figure}

XML stands for Extensible Markup Language.
It is a metalanguage that allows anyone to define their own customized markup languages.
To astronomers, XML might be most familiar due its similarity to HTML, the language in which the World Wide Web is written.
However, XML is not about the displaying or formatting but rather about the creating and sharing of information.
A document is a tree structure that consists of tags such as
\begin{lstlisting}[language=XML]
<tag attributename="attributevalue">text</tag>.
\end{lstlisting}
Each tag can contain text and/or other tags.
Tags can also contain optional attributes.
In contrast to HTML, each tag must be closed explicitly. 
XML contains a lot more features such as the possibility to precisely define and verify a customized language.
Most of these features are not important for the following discussion and I will therefore not go into any more detail.

Listing \ref{lst:corot} shows an example of a database entry written in the XML file format. 
It describes the planetary system CoRoT-5 and is taken from the Open Exoplanet Catalogue.
Without any further introduction, header file or manual, any astronomer can pick out the most important pieces of information.

For example, you can easily convince yourself that this is a Hot Jupiter orbiting a 1.03 solar mass star at 0.04947 astronomical units and a period of 4.03 days. 
It is located 400 parsecs away from the Sun at the coordinates $\alpha=6^h45^m7^s, \delta=0^\circ48'55"$.
In this particular example there are no units given the in the XML file itself because of well established conventions (e.g. quoting the star's mass in units of Solar mass). 
However, adding units to tags in the form of an attribute is straightforward:
\begin{lstlisting}[language=XML]
<mass unit="solarmass">1.03</mass>.
\end{lstlisting}
Errorbars and any additional meta data can be added in a similar way.
Tags for information that does not exist for a given object can simply be removed.
This dramatically enhanced readability and also saves disk space.
One can also emphasize the missing value with an empty tag, for example:
\begin{lstlisting}[language=XML]
<temperature />.
\end{lstlisting}
I hope this example has convinced you that this format makes the entire database human readable.
This is of course no substitute for a full description of each fields and all the implicit conventions used.
But it takes away a lot of unnecessary overhead.

The XML file in Listing \ref{lst:corot} is also a good example to show how hierarchical information can be stored. 
The planet clearly belongs to the star.
Now, imagine someone discoveries a wide binary companion to CoRoT-5. 
Adding another star to the system is trivial, it is just another \xml{star} tag at the end of the file.
Even something exotic as adding a moon which is orbiting the planet can be added very naturally.
None of this breaks the structure of the database.
Scripts parsing this database remain fully functional, as described below\footnote{This statement is of course without loss of generality; one can always construct a script which will not work anymore.}.

Using only small text files to store the database has another big advantage over more traditional approaches that store everything in one big file or database such as MySQL: small text files work nicely together with version control systems. 
These systems were originally designed to track changes in source code.
It also makes sense form a logical point of view as each line contains exactly one piece of information.

\subsection{Scriptablility}
XML files are great for scripting.
Listing \ref{lst:python} shows an example of a short python script that queries the Open Exoplanet Catalogue and outputs the name and mass of all planets.
\begin{figure}
\caption{Example of a python script that outputs a list of planet names and masses from the Open Exoplanet Catalogue.}
\vspace{5pt}
\begin{lstlisting}[language=python, numbers=left,frame=tb]
import xml.etree.ElementTree as ET, glob
for filename in glob.glob("*.xml"):
	tree = ET.parse(open(filename, 'r'))
	planets = tree.findall(".//planet")
	for planet in planets:
		print planet.findtext("./name"), 
		print planet.findtext("./mass")
\end{lstlisting}
\vspace{5pt}
\label{lst:python}
\end{figure}

The use of a scripting language like python and the XML file format allows one to easily create complex queries. 
For example, imagine you are only interested in planets that are in multi-planetary system. 
Adding the single line 
\begin{lstlisting}[language=python]
if len(planets)>1:
\end{lstlisting}
after line 4 in Listing \ref{lst:python} is all that is needed to accomplish that.
The syntax is so simple that it does not even require any explanation.
Scripts can also be used to perform automatic updates to the database or even create the entire database from data in a completely different format.
Because the entire database is stored in a version control system, there is no risk of accidentally deleting information.

\section{Version control system \texttt{git}} \label{sec:git}
The version control system \texttt{git} has seen a tremendous success in the open source world.
It has become the de facto standard for distribution version control systems.
Several alternatives with almost equal functionality exist (for example \texttt{Bazaar} and \texttt{Mercurial}).
Although the following discussion is very general, it will focus on \git.
This is not meant to be an introduction to \texttt{git}.
I want to point out how \git{} can be used for maintaining an astronomical database.
There are many good tutorials on \texttt{git} and you are strongly encouraged to have a look at those.
An interactive and well made introduction can be found at \url{http://try.github.com/}.

\subsection{Workflow}
Every copy (fork) of a \git{} repository contains the entire history of every change ever committed to the database.
There is no central server, no master and no slaves. 
The copy on your laptop is completely equal to the copy hosted on your web-server.
However, more people might have access to the copy on your web-server.

When starting to work on a new database or catalogue, you create a new repository on a local machine and start adding data.
You can then push the repository to another publicly accessible or private computer  (ssh access is enough to allow \git{} to communicate). 
A possibility worth looking into is the online platform \url{http//github.com} that was already mentioned above.
It is a free hosting service for \git{} repositories.
Collaborators can then copy (pull) the repository and make their own changes and additions.
When they think their changes are good enough to require your attention, they will send you a pull request. 
You review the changes and check for consistency of the data files.
For example, you might need to manually resolve a conflict if the same line in the same file was edited by more than one person. 
Once the changes are approved, anyone else will be able to download the newest version.

There are several option when the maintainer does not agree or is not satisfied with the submitted changes.
He can simply ignore the changes or try to improve them.
He can also create a new branch within the same repository and call it \textit{experimental} or \textit{controversial}.
This allows others to see and discuss the submission but does not pollute the main database with questionable data.

It is worth noting that a human readable format uses more disk space than a binary format.
However, \texttt{git} automatically compresses data so that this should not be a big issue for most small to medium sized catalogues.
In addition to that, there is no need to add tags to the XML file which do not contain any data. 
For example, a tag containing the value for the Rossiter-McLaughlin effect \citep{Rossiter1924,McLaughlin1924} only needs to be added to those planets that had such an observation.

\subsection{Use of \git{}-based catalogues in other \git{} repositories}
\git{}'s submodule functionality is ideally suited for databases and catalogues.
It allows an external git repository to be added to another git repository as a submodule.
This creates a logical separation between a database and scripts that make use of it. 
But at the same time, this configuration can keep the database in sync with the public version.

\subsection{References to scientific publications}
Including references is absolutely essential for any scientific database. 
However, it increases the size, readability and complexity of any data format dramatically.
Once again, \git{} comes to our rescue. 
The by far simplest way to add reference to a database that is storred in \git{} is the commit message. 
This enables anyone to trace back each value to the person who made the entry in the database, but also to the scientific publication where the information is coming from.
Note that this is only possible if the file format is constructed such that there is one value (or at most one set of values that comes from the same publication) per line.

\section{The Open Exoplanet Catalogue}
The Open Exoplanet Catalogue is an example of a database that incorporates all the ideas presented above. 
It is a daily updated catalogue of all discovered extra-solar planets.
Furthermore, it is to my knowledge the only catalogue that correctly stores and represents planets in an arbitrary star system (s- and p-type binaries, triple and quadruple star systems, etc).
The catalogue is hosted on github at \url{https://github.com/hannorein/open\_exoplanet\_catalogue/}.

The project was started in late 2010. 
The motivation came from the simple fact that I needed a catalogue of all discovered extra-solar planets but I was unhappy with existing catalogues.
Some catalogues, for example \url{http://exoplanets.org} \citep{Wright2011} are updated only irregularly and often lag several weeks behind important discoveries.
Other websites, most importantly \url{http://exoplanet.eu} \citep{Schneider2011}, are usually updated very quickly after an announcement is made.
But there were several typos and other inconsistencies in the catalogue that I needed to keep track of and change manually after every update because there was no way for me to directly contribute to the database itself.
In addition to that, certain pieces of information (such as the discovery method) were available on the website, but not in the downloadable files.
I ended up having to parse the (always changing) HTML files to find that information. 

The original dataset was taken from the two websites mentioned above but has since been modified heavily. 
As of December 4th 2012 the database contains 861 planets.
The catalogue also contains a separate dataset of the latest Kepler Objects of Interest \citep{Borucki2011b} which uses the same data format.
The compressed size of the main catalogue without images is less than 80kb.
It is also used as an input catalogue for the iOS application Exoplanet\footnote{\label{noteapp}\url{http://exoplanetapp.com}}.


\subsection{Planets in multiple star system}
The Open Exoplanet Catalogue can represent planetary systems in arbitrary hierarchical systems. 
The basic structure contains only three objects: \texttt{<planet>}, \texttt{<star>} and \texttt{<binary>}.
A binary hosts two stars, a star and a binary or two other binaries.
Both stars and binaries can host planets.
These simple rules allows one to construct any system. 
Listing \ref{lst:ph1} shows the circumbinary system PH-1 (AaAb) b \cite[also known as Kepler-64, see][]{Schwamb2012} which is part of a quadruple system.

\begin{figure}
\caption{Example of the database entry of the circumbinary planet PH-1b from the Open Exoplanet Catalogue.}
\vspace{5pt}
\begin{lstlisting}[language=XML, numbers=left,frame=tb]
<name>PH-1</name>
<binary>
	<semimajoraxis>1000</semimajoraxis>
	<binary>
		<semimajoraxis>0.1744</semimajoraxis>
		<star>
			<name>PH-1 Aa</name>
			<mass>1.384</mass>
		</star>
		<star>
			<name>PH-1 Ab</name>
			<mass>0.386</mass>
		</star>
		<planet>
			<name>PH-1 (AaAb) b</name>
			<name>Kepler-64 b</name>
			<semimajoraxis>0.634&\newline&</semimajoraxis>
		</planet>
	</binary>
	<binary>
		<semimajoraxis>60</semimajoraxis>
		<star>
			<name>PH-1 Ba</name>
			<mass>0.99</mass>
		</star>
		<star>
			<name>PH-1 Bb</name>
			<mass>0.51</mass>
		</star>
	</binary>
</binary>
\end{lstlisting}
\vspace{5pt}
\label{lst:ph1}
\end{figure}

In a standard, non-hierarchical database such as a simple table, it is almost impossible to represent systems like these.
On the other hand the flexibility of a tree based file format like XML, allows one to add these on the fly to an existing database and in a very natural way. 
Note that the script in Listing \ref{lst:python} is still fully functional even for this complicated quadruple system.

\newpage
\subsection{Images and other binary data}
The Open Exoplanet Catalogue contains image files of directly imaged exoplanets and artistic illustrations often released by various space agencies.
This is an example of how binary data and images can be attached to a catalogue based on the XML file format.
Binary data can be included in XML files in the same way e-mails can include attachments\footnote{For example, this can be done using a so called Multipart/related MIME type, RFC 2112, see \url{http://tools.ietf.org/html/rfc2112}.}.
In most cases a much simpler approach, storing images and binary data in a separate file, is sufficient.
To link the XML file to the binary file, one only has to add a tag with the filename.

In the Open Exoplanet Catalogue, the \xml{image} tag is used to store images related to a certain planet. 
An additional tag \xml{imagedescription} tag is used to store a description, a reference to the scientific paper and copyright information for the image.

\subsection{Other user interfaces}
The entire database can be directly accessed via github. 
Because each entry is a human-readable XML file, one can simply read that file online in a web browser.
In fact, some of these files might show up on the first page when you google a specific planetary system.

However, an interactive and graphical user interface is often very useful.
I created such a website, the \textit{Visual Exoplanet Catalogue}\footnote{\url{http://exoplanet.hanno-rein.de}}.
It is based on the Open Exoplanet Catalogue and is using it as a back-end.
The catalogue is also easily accessible with the free iOS application \textit{Exoplanet}$^\text{\ref{noteapp}}$.
I'm looking forward to other new and exciting projects that will make use of this catalogue.

\subsection{List of tags}
Table \ref{tab:tags} contains a list and the definitions of all tags currently used in the Open Exoplanet Catalogue. 
This list will certainly grow with time.

\begin{table*}
\centering
\begin{tabular}{>{\raggedright}p{2.7cm}|>{\raggedright}p{2.7cm}|>{\raggedright}p{7.5cm}|p{3.2cm}}
{\scshape Tag}& {\scshape Can be child of} & {\scshape Description} & {\scshape Unit}\\ \hline\hline
\xml{system}		&  - &This is the root container for an entire planetary system & \\
\xml{planet} 		& \xml{system}, \xml{binary}, \xml{star} & This is the container for a single planet. The planet is a free floating (orphan) planet if this tag is a child of \xml{system}. & \\
\xml{star} 		& \xml{system}, \xml{binary} & This is the container for a single star. A star can be host to one or more planets (circum-stellar planets). &\\
\xml{binary} 		& \xml{system}, \xml{binary} & A binary consists of either two stars, one star and one binary or two binaries. In addition a binary can be host to one or more planets (circum-binary planets).&\\
\hline
\xml{declination}	& \xml{system} & Declination & $\pm$dd mm ss  \\
\xml{rightascension}	& \xml{system} & Right ascension & hh mm ss  \\
\xml{distance}		& \xml{system} & Distance from the Sun & parsec  \\
\xml{name}		& \xml{system}, \xml{star}, \xml{planet} & Name of this object. This tag can be used multiple times if the object has multiple Names. &  \\
\xml{semimajoraxis} 	& \xml{binary}, \xml{planet} & Semi-major axis of a planet (heliocentric coordinates) if child of \xml{planet}. Semi-major axis of the binary if child of \xml{binary}. &  AU\\
\xml{eccentricity} 	& \xml{binary}, \xml{planet} & Eccentricity  &\\
\xml{periastron} 	& \xml{binary}, \xml{planet} & Longitude of periastron & degree \\
\xml{longitude} 	& \xml{binary}, \xml{planet} & Mean longitude at a given Epoch (same for all planets in one system) & degree \\
\xml{ascendingnode} 	& \xml{binary}, \xml{planet} & Longitude of the ascending node & degree \\
\xml{inclination} 	& \xml{binary}, \xml{planet} & Inclination of the orbit & degree \\
\xml{period}	 	& \xml{binary}, \xml{planet} & Orbital period   & day \\
\xml{mass}		& \xml{planet}, \xml{star} &Mass (or $m \sin(i)$ for radial velocity planets) & Jupiter masses (\xml{planet}), Solar masses (\xml{star}) \\
\xml{radius}		& \xml{planet}, \xml{star} &Physical radius & Jupiter radii (\xml{planet}), Solar radii (\xml{star}) \\
\xml{temperature}	& \xml{planet}, \xml{star} &Temperature (surface or equilibrium) & Kelvin \\
\xml{age}		& \xml{planet}, \xml{star} &Age & Gyr \\
\xml{metallicity}	& \xml{star} & Stellar metallicity  & log, relative to solar \\
\xml{spectraltype}	& \xml{star} & Spectral type  &  \\
\xml{magV}		& \xml{star} & Visual magnitude &  \\
\hline
\xml{discoverymethod} 	& \xml{planet} & Discovery method of the planet. For example: timing, RV, transit, imaging.  &  \\
\xml{description} 	& \xml{planet} & Short description of the planet  &  \\
\xml{image} 		& \xml{planet} & Filename without extension of a picture of the planet. File is stored in the images directory. &  \\
\xml{imagedescription}	& \xml{planet} & Short description and copyright information of the image. &  \\
\xml{new}		& \xml{planet} & The value for this tag is 1 if the system has been added recently (since the last update or within 48 hours) &  \\
\xml{discoveryyear}	& \xml{planet} & Year of the planet's discovery & yyyy \\
\xml{lastupdate}	& \xml{planet} & Date of the last (non-trivial) update & yy/mm/dd  \\
\end{tabular}
\caption{All tags currently used in the Open Exoplanet Catalogue.}
\label{tab:tags}
\end{table*}

\section{Conclusions}
My experience of talking to astronomers and astrophysicists is that many are not very familiar with the resources and tools that the computer science and open source community uses.
However, the benefits listed in Section~\ref{sec:adv} which these tools can provide to any new catalogue or database are tremendous. 
A long ASCII table, a FITS file, a MySQL database or a complicated web-form is just not good enough anymore.
With this proposal and the Open Exoplanet Catalogue, I hope to encourage discussion, experimentation and innovation.

The ideas and concepts presented here are not meant to be final and leave a lot of room for improvements.
Many of the details will strongly depend on the specific kind and usage scenario of the database.
The examples here are biased towards a catalogue of extra-solar planets.
Other areas where heterogeneous datasets are compiled into catalogues and where ideas very similar to those presented here can be readily applied are for example a catalogue of gravitational lenses and a database of solar system objects.

The Open Exoplanet Catalogue has already proven to be very useful for several of my own projects \citep[e.g.][]{Rein2012b}. 
But it is by no means perfect. 
For example, it is missing important information such as the Rossiter-McLaughlin measurements and most certainly contains errors.
But the important point is that YOU can help improve it.
Visit the catalogue's website on github, create an account, clone the repository and if you find a typo or want to add new data, simply make the changes (there is no need to ask for permission).
If you are an observer, please consider adding your own data when you announce a new discovery.
The idea of a democratic, decentralized astronomical database can only work if many people contribute to it.
I am excited to find out if we, as a community, can find enough people to maintain this catalogue.  

If you would like to contribute to the catalogue but have problems getting started with \git{} or if you have any comments on this proposal, please do not hesitate to contact me.

\section*{Acknowledgments}
Hanno Rein would like to thank Dave Spiegel for inspiring conversations regarding the structure of the Open Exoplanet Catalogue.
Scott Tremaine, Dave Spiegel, Chi-kwan Chan, Leonidas Moustakas, Peter Williams and Wladimir Lyra gave valuable feedback on this paper.
Hanno Rein was supported by the Institute for Advanced Study and the NFS grant AST-0807444.
Support for the Exoplanet App came from the Royal Astronomical Society via a public outreach grant.

\bibliography{full}
\end{document}